\documentclass[aip,reprint]{revtex4-2}
\usepackage[T1]{fontenc}
\usepackage{times}
\usepackage{graphicx,color}
\usepackage{amsfonts,amsmath,amssymb,amsbsy}
\usepackage[colorlinks=true,linkcolor=blue,citecolor=blue,urlcolor=blue]{hyperref}
\def\H{{\mathcal{H}}}
\let\mathbf=\boldsymbol
\def\blue#1{\textcolor{blue}{#1}}

\def\blue#1{\textcolor{black}{#1}}
\def\emph#1{\textcolor{magenta}{#1}}
\begin{document}
%%%%%%%%%%%%%%%%%%%%%%%%%%%%%%%%%%%%%%%%%%%%%%%%%%%%%%%%%%%%

\title{A Frustrated Bimeronium: Static Structure and Dynamics}

\author{Xichao Zhang}
%\email[Email:~]{xichaozhang@gmail.com}
\thanks{X.Z. and J.X. contributed equally to this work.}
\affiliation{School of Science and Engineering, The Chinese University of Hong Kong, Shenzhen, Guangdong 518172, China}

\author{Jing Xia}
%\email[Email:~]{jingxia0817@gmail.com}
\thanks{X.Z. and J.X. contributed equally to this work.}
\affiliation{College of Physics and Electronic Engineering, Sichuan Normal University, Chengdu 610068, China}

\author{Motohiko Ezawa}
\thanks{Authors to whom correspondence should be addressed:~ezawa@ap.t.u-tokyo.ac.jp~and~zhouyan@cuhk.edu.cn}
%\email[Email:~]{ezawa@ap.t.u-tokyo.ac.jp}
\affiliation{Department of Applied Physics, The University of Tokyo, 7-3-1 Hongo, Tokyo 113-8656, Japan}

\author{Oleg A. Tretiakov}
%\email[Email:~]{o.tretiakov@unsw.edu.au}
\affiliation{School of Physics, The University of New South Wales, Sydney 2052, Australia}
\affiliation{National University of Science and Technology ``MISiS'', Moscow 119049, Russia}

\author{Hung T. Diep}
%\email[Email:~]{diep@u-cergy.fr}
\affiliation{Laboratoire de Physique Th{\'e}orique et Mod{\'e}lisation, Universit{\'e} de Cergy-Pontoise, 95302 Cergy-Pontoise Cedex, France}

\author{Guoping Zhao}
%\email[Email:~]{zhaogp@uestc.edu.cn}
\affiliation{College of Physics and Electronic Engineering, Sichuan Normal University, Chengdu 610068, China}

\author{Xiaoxi Liu}
%\email[Email:~]{liu@cs.shinshu-u.ac.jp}
\affiliation{Department of Electrical and Computer Engineering, Shinshu University, 4-17-1 Wakasato, Nagano 380-8553, Japan}

\author{Yan Zhou}
\thanks{Authors to whom correspondence should be addressed:~ezawa@ap.t.u-tokyo.ac.jp~and~zhouyan@cuhk.edu.cn}
%\email[Email:~]{zhouyan@cuhk.edu.cn}
\affiliation{School of Science and Engineering, The Chinese University of Hong Kong, Shenzhen, Guangdong 518172, China}

%-%-%-%-%-%-%-%-%-%-%-%-%-%-%-%-%-%-%-%-%-%-%-%-%-%-%-%-%-%-%
\begin{abstract}
We show a topological spin texture called ``bimeronium'' in magnets with in-plane magnetization. It is a topological counterpart of skyrmionium in perpendicularly magnetized magnets and can be seen as a combination of two bimerons with opposite topological charges. We report the static structure and spin-orbit-torque-induced dynamics of an isolated bimeronium in a magnetic monolayer with frustrated exchange interactions. We study the anisotropy and magnetic field dependences of a static bimeronium. We also explore the bimeronium dynamics driven by the damping-like spin-orbit torque. We find that the bimeronium shows steady rotation when the spin polarization direction is parallel to the easy axis. Moreover, we demonstrate the annihilation of the bimeronium when the spin polarization direction is perpendicular to the easy axis. Our results are useful for understanding fundamental properties of bimeronium structures and may offer an approach to build bimeronium-based spintronic devices.
\end{abstract}
%-%-%-%-%-%-%-%-%-%-%-%-%-%-%-%-%-%-%-%-%-%-%-%-%-%-%-%-%-%-%

%\date{\today}
%\date{2 May 2020}
\date{21 January 2021}

%\preprint{\textsl{\textcolor[rgb]{0.5,0.5,0.5}{Re-submitted to Applied Physics Letters (Special Topic on Mesoscopic Magnetic Systems: From Fundamentals Properties to Devices)}}}

%\keywords{skyrmion, skyrmionium, bimeron, bimeronium, frustrated magnet, spintronics, micromagnetics}
%\pacs{75.10.Hk, 75.10.Jm, 75.70.Ak, 75.70.Kw, 75.78.-n, 12.39.Dc}
% Classical spin models:                                     75.10.Hk
% Quantized spin models, including quantum spin frustration: 75.10.Jm
% Magnetic properties of monolayers and thin films:          75.70.Ak
% Domain structure (magnetic bubbles and vortices):          75.70.Kw
% Magnetization dynamics:                                    75.78.-n
% Skyrmions:                                                 12.39.Dc

\maketitle

%-%-%-%-%-%-%-%-%-%-%-%-%-%-%-%-%-%-%-%-%-%-%-%-%-%-%-%-%-%-%
%\section{Introduction}
%\label{se:Introduction}
%-%-%-%-%-%-%-%-%-%-%-%-%-%-%-%-%-%-%-%-%-%-%-%-%-%-%-%-%-%-%

Topological magnetism and spin frustration are important and hot topics in the fields of magnetism and spintronics~\cite{Bogdanov_1989,Roszler_NATURE2006,Nagaosa_NNANO2013,Finocchio_JPD2016,Kang_PIEEE2016,Kanazawa_AM2017,Wanjun_PHYSREP2017,Fert_NATREVMAT2017,Zhou_NSR2018,ES_JAP2018,Zhang_JPCM2020,Diep_Entropy2019,Gobel_PR2020}.
The link between topological magnetism and spin frustration lies in the fact that many topological spin textures can be stabilized in frustrated spin systems~\cite{Okubo_PRL2012,Leonov_NCOMMS2015,Lin_PRB2016A,Hayami_PRB2016A,Rozsa_PRL2016,Leonov_NCOMMS2017,Kharkov_PRL2017,Xichao_NCOMMS2017,Yuan_PRB2017,Hu_SR2017,Malottki_SR2017,Hou_AM2017,Liang_NJP2018,Ritzmann_NE2018,Xia_PRApplied2019,Kurumaji_SCIENCE2019,Desplat_PRB2019,Zarzuela_PRB2019,Lohani_PRX2019,Gobel_PRB2019,Diep_PRB2019,Diep_Symmetry2020,Zhang_PRB2020,Xia_APL2020}.
For example, the magnetic skyrmion is an exemplary topological spin texture~\cite{Bogdanov_1989,Roszler_NATURE2006}, which can be regarded as a quasi-particle and shows intriguing dynamics~\cite{Nagaosa_NNANO2013,Finocchio_JPD2016,Kang_PIEEE2016,Kanazawa_AM2017,Wanjun_PHYSREP2017,Fert_NATREVMAT2017,Zhou_NSR2018,ES_JAP2018,Zhang_JPCM2020,Diep_Entropy2019,Gobel_PR2020}.

The magnetism and spintronics community has focused on skyrmions stabilized by the Dzyaloshinskii-Moriya (DM) interaction~\cite{Muhlbauer_SCIENCE2009,Yu_NATURE2010,Woo_NMATER2016,Wanjun_NPHYS2017,Litzius_NPHYS2017,Rosales_2015,Rosales_2020}, however, recent progress in the field revealed that skyrmions and other topological spin textures can be found in a different system, where topological spin textures are stabilized by exchange frustration~\cite{Okubo_PRL2012,Leonov_NCOMMS2015,Lin_PRB2016A,Hayami_PRB2016A,Rozsa_PRL2016,Leonov_NCOMMS2017,Kharkov_PRL2017,Xichao_NCOMMS2017,Yuan_PRB2017,Hu_SR2017,Malottki_SR2017,Hou_AM2017,Liang_NJP2018,Ritzmann_NE2018,Xia_PRApplied2019,Kurumaji_SCIENCE2019,Desplat_PRB2019,Zarzuela_PRB2019,Lohani_PRX2019,Gobel_PRB2019,Diep_PRB2019,Diep_Symmetry2020,Zhang_PRB2020,Xia_APL2020}.
Typical frustrated topological spin textures include the skyrmion~\cite{Okubo_PRL2012,Leonov_NCOMMS2015,Lin_PRB2016A,Hayami_PRB2016A,Rozsa_PRL2016,Leonov_NCOMMS2017,Kharkov_PRL2017,Xichao_NCOMMS2017,Yuan_PRB2017,Hu_SR2017,Malottki_SR2017,Hou_AM2017,Liang_NJP2018,Ritzmann_NE2018,Xia_PRApplied2019,Kurumaji_SCIENCE2019,Desplat_PRB2019,Zarzuela_PRB2019,Lohani_PRX2019,Gobel_PRB2019,Diep_PRB2019,Diep_Symmetry2020,Zhang_PRB2020},
the skyrmionium~\cite{Xia_APL2020,Komineas_PRB2015A,Komineas_PRB2015B} (i.e., target skyrmion~\cite{Rozsa_PRB2018} and $2\pi$-skyrmion~\cite{Zheng_PRL2017}),
and the bimeron~\cite{Zhang_PRB2020,Kharkov_PRL2017,Gobel_PRB2019} (i.e., asymmetric skyrmion~\cite{Leonov_PRB2017} and meron pair~\cite{Lin_PRB2015}).
Indeed, skyrmioniums and bimerons can also be stabilized by the DM interaction~\cite{Bogdanov_JMMM1999,Finazzi_PRL2013,Beg_SREP2015,Xichao_PRB2016,Zhang_NanoLett2018,Gobel_PRB2019,Kuchkin_PRB2020}.
In principle, all of these particle-like topological spin textures can be used to carry information~\cite{Finocchio_JPD2016,Kang_PIEEE2016,Wanjun_PHYSREP2017,Fert_NATREVMAT2017,Zhou_NSR2018,ES_JAP2018,Zhang_JPCM2020,Gobel_PR2020}, and thus are promising for building future information storage and logic computing devices~\cite{Finocchio_JPD2016,Kang_PIEEE2016,Wanjun_PHYSREP2017,Fert_NATREVMAT2017,Zhou_NSR2018,ES_JAP2018,Zhang_JPCM2020,Gobel_PR2020}.

\begin{figure}[t]
\centerline{\includegraphics[width=0.485\textwidth]{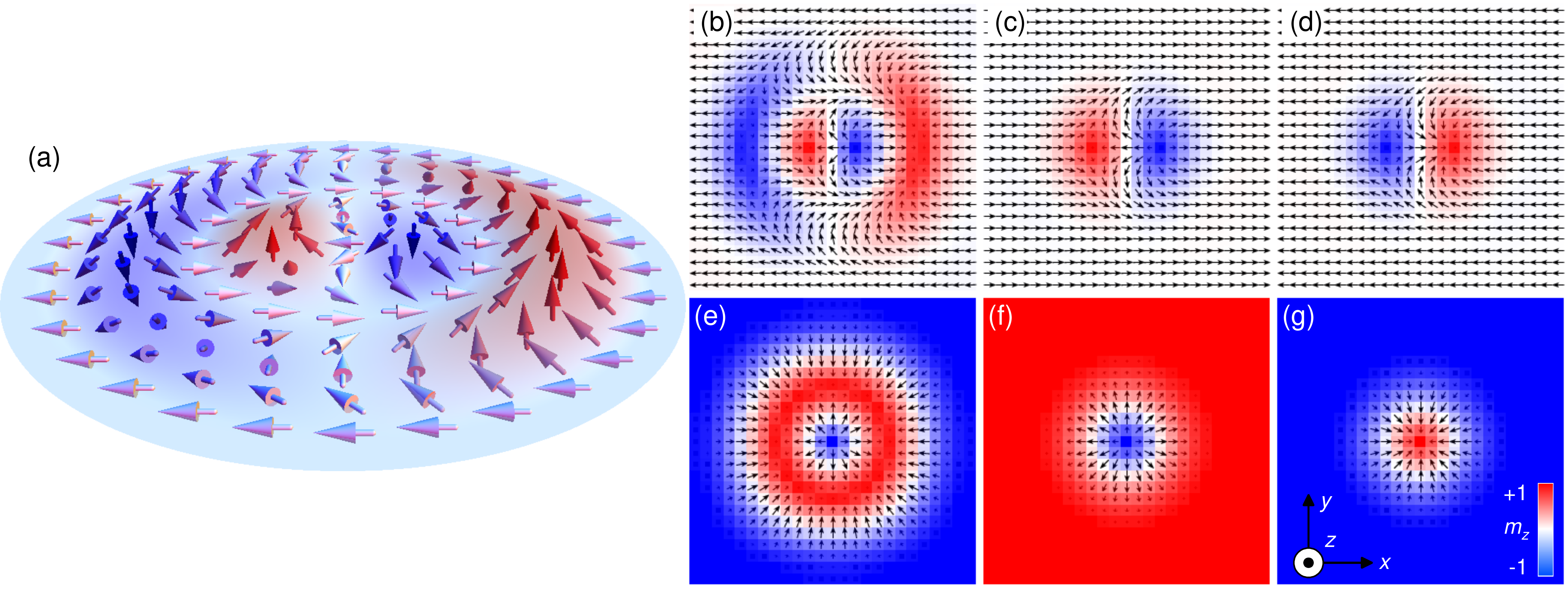}}
\caption{%
(a) Schematic illustration of a bimeronium with $Q=0$. The arrows represent the spin directions. The out-of-plane spin component ($m_z$) is color coded.
(b) Top view of a relaxed bimeronium with $Q=0$.
(c) Top view of a relaxed bimeron with $Q=-1$.
(d) Top view of a relaxed bimeron with $Q=+1$.
(e) Top view of a relaxed skyrmionium with $Q=0$.
(f) Top view of a relaxed skyrmion with $Q=-1$.
(g) Top view of a relaxed skyrmion with $Q=+1$.
Here, $J_2=-0.8$, $J_3=-0.9$, $K=0.02$, and $B=0$. The displayed area is of $10\times 10$ nm$^{2}$. For [(b)-(d)], the easy axis is aligned along the $x$ direction. For [(e)-(g)], the easy axis is aligned along the $z$ direction. The displayed area is of $10\times 10$ nm$^{2}$.
}
\label{FIG1}
\end{figure}

In this Letter, we report that the topological counterpart of skyrmioniums, which is called the \textit{bimeronium} [Fig.~\ref{FIG1}(a)], can be stabilized in an in-plane magnetic system with competing exchange interactions.
We study the static structure of an isolated bimeronium with a topological charge of zero at different anisotropies and magnetic fields.
We also investigate the dynamics of an isolated bimeronium induced by the damping-like spin-orbit torque (SOT).
Our results suggest that the frustrated bimeronium could be used as a special building block for spintronic applications, however, it cannot move like the frustrated bimeron~\cite{Zhang_PRB2020} due to its complex and non-circular symmetric spin structure that could be annihilated by the SOT at certain conditions.

%-%-%-%-%-%-%-%-%-%-%-%-%-%-%-%-%-%-%-%-%-%-%-%-%-%-%-%-%-%-%
%\section{Methods}
%\label{se:Methods}
%-%-%-%-%-%-%-%-%-%-%-%-%-%-%-%-%-%-%-%-%-%-%-%-%-%-%-%-%-%-%

Our simulated system is a $J_{1}$-$J_{2}$-$J_{3}$ classical Heisenberg model on a simple monolayer square lattice~\cite{Kaul_2004,Lin_PRB2016A,Xichao_NCOMMS2017,Xia_PRApplied2019,Zhang_PRB2020,Xia_APL2020,Diep_Entropy2019}, where the nearest-neighbor (NN) exchange interaction $J_1$ is ferromagnetic (FM), while the next-NN (NNN) $J_2$ and next-NNN (NNNN) $J_3$ exchange interactions are antiferromagnetic (AFM).
The Hamiltonian $\mathcal{H}$ includes the FM and AFM exchange interactions, in-plane easy-axis magnetic anisotropy ($K$), and applied magnetic field ($\boldsymbol{B}$), given as
\begin{align}
\H=&-J_1\sum_{\substack{<i,j>}}\boldsymbol{m}_i\cdot\boldsymbol{m}_j-J_2\sum_{\substack{\ll i,j\gg}}\boldsymbol{m}_i\cdot\boldsymbol{m}_j \\
&-J_3\sum_{\substack{\lll i,j\ggg}}\boldsymbol{m}_i\cdot\boldsymbol{m}_j-K\sum_{\substack{i}}{(m^x_i)^{2}}-\sum_{\substack{i}}{\boldsymbol{B}\cdot\boldsymbol{m}_i}, \notag 
\label{eq:Hamiltonian}
\end{align}
where $\boldsymbol{m}_{i}$ represents the normalized spin at the site $i$, $|\boldsymbol{m}_{i}|=1$.
$\left\langle i,j\right\rangle$,
$\left\langle\left\langle i,j\right\rangle\right\rangle$, and
$\left\langle\left\langle\left\langle i,j\right\rangle\right\rangle\right\rangle$
run over all the NN, NNN, and NNNN sites in the magnetic monolayer, respectively.
$K$ is the easy-axis magnetic anisotropy constant, and the easy axis direction is aligned along the $x$ axis.
$\boldsymbol{B}$ is the applied magnetic field.

The spin dynamics is simulated by using the Object Oriented MicroMagnetic Framework (OOMMF)~\cite{OOMMF} with our extension modules for the $J_{1}$-$J_{2}$-$J_{3}$ classical Heisenberg model~\cite{Lin_PRB2016A,Xichao_NCOMMS2017,Xia_PRApplied2019,Zhang_PRB2020,Xia_APL2020}.
We also use the OOMMF conjugate gradient minimizer for obtaining relaxed spin structures, which is a method that locates local minima in the energy surface through direct minimization techniques~\cite{OOMMF}.
The spin dynamics is governed by the Landau-Lifshitz-Gilbert (LLG) equation~\cite{OOMMF}
\begin{equation}
\frac{d\boldsymbol{m}}{dt}=-\gamma_{0}\boldsymbol{m}\times\boldsymbol{h}_{\rm{eff}}+\alpha\left(\boldsymbol{m}\times\frac{d\boldsymbol{m}}{dt}\right),
\label{eq:LLG}
\end{equation}
where $|\boldsymbol{m}|=1$ represents the normalized spin,
$\boldsymbol{h}_{\rm{eff}}=-\frac{1}{\mu_{0}M_{\text{S}}}\cdot\frac{\delta\mathcal{H}}{\delta\boldsymbol{m}}$ is the effective field,
$t$ is the time,
$\alpha$ is the Gilbert damping parameter,
$\gamma_0$ is the absolute value of the gyromagnetic ratio,
and $M_{\text{S}}$ is the saturation magnetization.

For the spin dynamics driven by the SOT, we consider the damping-like SOT $\tau_{\text{d}}$ expressed as~\cite{Finocchio_JPD2016,ES_JAP2018,Zhang_JPCM2020,Tomasello_SREP2014}
$\tau_{\text{d}}=\frac{u}{a}\left(\boldsymbol{m}\times\boldsymbol{p}\times\boldsymbol{m}\right)$,
where $u=\left|\left(\gamma_{0}\hbar/\mu_{0}e\right)\right|\cdot\left(j\theta_{\text{SH}}/2 M_{\text{S}}\right)$ is the spin torque coefficient.
$\hbar$ is the reduced Planck constant, $e$ is the electron charge, $\mu_{0}$ is the vacuum permeability constant, $a$ is the thickness of the FM monolayer (i.e., the lattice constant here),
$j$ is the applied current density, $\theta_{\text{SH}}$ is the spin Hall angle.
$\boldsymbol{p}$ denotes the spin polarization direction.
$\tau_\text{d}$ is added to the right-hand side of Eq.~(\ref{eq:LLG}) when the damping-like SOT is turned on.

In this work, we define the topological charge $Q$ in the continuum limit by the formula~\cite{Nagaosa_NNANO2013,Zhang_JPCM2020}
$Q={\frac{1}{4\pi}}\int\boldsymbol{m}(\boldsymbol{r})\cdot\left(\partial_{x}\boldsymbol{m}(\boldsymbol{r})\times\partial_{y}\boldsymbol{m}(\boldsymbol{r})\right)d^{2}\boldsymbol{r}$.
We parametrize the bimeronium [Fig.~\ref{FIG1}(b)] and bimeron [Fig.~\ref{FIG1}(c)-(d)] as
$\boldsymbol{m}(\boldsymbol{r})=\boldsymbol{m}(\theta,\phi)=(\cos\theta,\sin\theta\sin\phi,-\sin\theta\cos\phi)$,
and we parametrize the skyrmionium [Fig.~\ref{FIG1}(e)] and skyrmion [Fig.~\ref{FIG1}(f)-(g)] as
$\boldsymbol{m}(\boldsymbol{r})=\boldsymbol{m}(\theta,\phi)=(\sin\theta\cos\phi,\sin\theta\sin\phi,\cos\theta)$.
We define
$\phi=Q_{\text{v}}\varphi+\eta$,
where $\varphi$ is the azimuthal angle in the $y$-$z$ plane ($0\le\varphi<2\pi$).
For the bimeron and skyrmion, we assume that $\theta$ rotates $\pi$ for spins from the texture center to the texture edge~\cite{Bogdanov_JMMM1999}.
For the bimeronium and skyrmionium, we assume that $\theta$ rotates $2\pi$ for spins from the texture center to the texture edge~\cite{Bogdanov_JMMM1999}.
Hence, $Q_{\text{v}}=\frac{1}{2\pi}\oint_{C}d \phi$ is the vorticity and $\eta$ is the helicity defined mod $2\pi$. Note that $\eta=0$ is identical to $\eta=2\pi$.

The default simulation parameters are~\cite{Lin_PRB2016A,Xichao_NCOMMS2017,Xia_PRApplied2019,Zhang_PRB2020,Xia_APL2020}:
$J_1=30$ meV,
$J_2=-0.8$ (in units of $J_1=1$),
$J_3=-0.9$ (in units of $J_1=1$),
$K=0.02$ (in units of $J_{1}/a^{3}=1$),
$B=0$ (in units of $J_{1}/a^{3}M_{\text{S}}=1$),
$\alpha=0.3$,
$\gamma_0=2.211\times 10^{5}$ m A$^{-1}$ s$^{-1}$,
$\theta_{\text{SH}}=0.2$,
and $M_{\text{S}}=580$ kA m$^{-1}$.
The lattice constant is $a=0.4$ nm (i.e., the mesh size is $0.4 \times 0.4 \times 0.4$ nm$^3$).
We have simulated the metastability diagram showing that the frustrated bimeronium can be a metastable state for a wide range of $J_2$ and $J_3$ (see \blue{supplemental material}).
The minimum required value of $J_3$ for stabilizing bimeroniums decreases with increasing $J_2$.

%%%%%%%%%%%%%%%%%%%%%%%%%%%%%%%%%%%%%%%%%%%%%%%%%%%%%%%%%%%%
%%%%%%%%%%%%%%%%%%%%%%%%%%%%%%%%%%%%%%%%%%%%%%%%%%%%%%%%%%%%

%%%%%%%%%%%%%%%%%%%%%%%%%%%%%%%%%%%%%%%%%%%%%%%%%%%%%%%%%%%%
\begin{figure}[t]
\centerline{\includegraphics[width=0.479\textwidth]{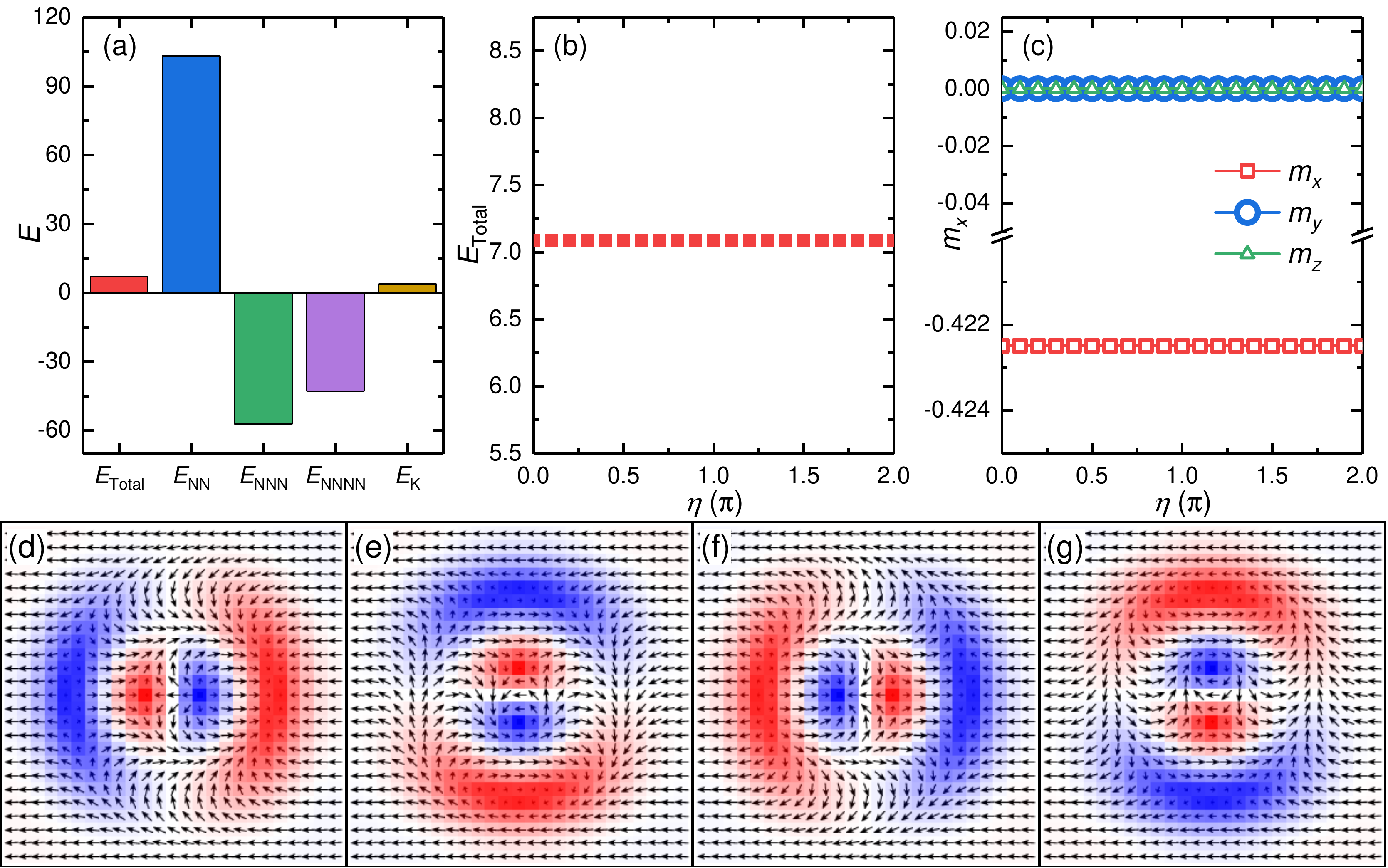}}
\caption{%
(a) Different energy contributions for a relaxed bimeronium with $\eta=0$. The energies are given in units of $J_1=1$.
(b) Total energy of a relaxed bimeronium as a function of $\eta$.
(c) Spin components of a relaxed bimeronium as functions of $\eta$.
Top views of relaxed bimeroniums with (d) $\eta=0$, (e) $\eta=\pi/2$, (f) $\eta=\pi$, and (g) $\eta=3\pi/2$ are given.
Here, $J_2=-0.8$, $J_3=-0.9$, $K=0.02$, and $B=0$. The arrows represent the spin directions. The out-of-plane spin component ($m_z$) is color coded. The displayed area is of $10\times 10$ nm$^{2}$.
}
\label{FIG2}
\end{figure}
%%%%%%%%%%%%%%%%%%%%%%%%%%%%%%%%%%%%%%%%%%%%%%%%%%%%%%%%%%%%

%%%%%%%%%%%%%%%%%%%%%%%%%%%%%%%%%%%%%%%%%%%%%%%%%%%%%%%%%%%%
\begin{figure}[t]
\centerline{\includegraphics[width=0.479\textwidth]{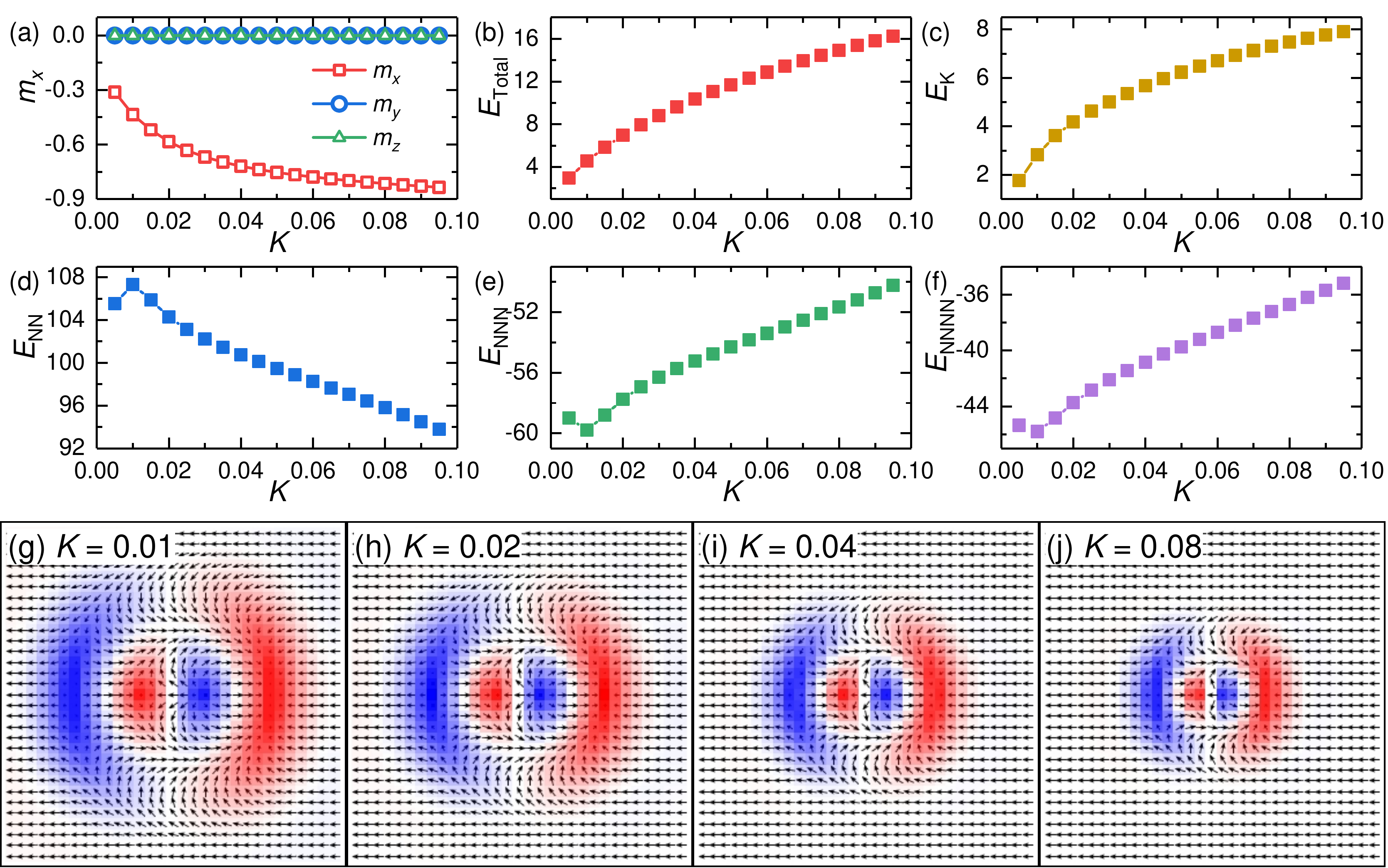}}
\caption{%
(a) Spin components as functions of $K$. A bimeronium with $\eta=0$ is relaxed at the center of a monolayer with $J_2=-0.8$, $J_3=-0.9$, and $B=0$.
(b) Total energy $E_{\text{Total}}$ as a function of $K$. The energies are given in units of $J_1=1$.
(c) Anisotropy energy $E_{\text{K}}$ as a function of $K$.
(d) NN exchange energy $E_{\text{NN}}$ as a function of $K$.
(e) NNN exchange energy $E_{\text{NNN}}$ as a function of $K$.
(f) NNNN exchange energy $E_{\text{NNNN}}$ as a function of $K$.
Top views of relaxed bimeroniums with $\eta=0$ at (g) $K=0.01$, (h) $K=0.02$, (i) $K=0.04$, and (j) $K=0.08$ are given.
The arrows represent the spin directions. The out-of-plane spin component ($m_z$) is color coded. The displayed area is of $12.4\times 12.4$ nm$^{2}$.
}
\label{FIG3}
\end{figure}
%%%%%%%%%%%%%%%%%%%%%%%%%%%%%%%%%%%%%%%%%%%%%%%%%%%%%%%%%%%%

%-%-%-%-%-%-%-%-%-%-%-%-%-%-%-%-%-%-%-%-%-%-%-%-%-%-%-%-%-%-%
%\section{Results and Discussion}
%\label{se:Results}
%-%-%-%-%-%-%-%-%-%-%-%-%-%-%-%-%-%-%-%-%-%-%-%-%-%-%-%-%-%-%

We first study the static structures and properties of a relaxed isolated bimeronium in the magnetic monolayer with competing exchange interactions and in-plane easy-axis anisotropy, where we set $J_2=-0.8$, $J_3=-0.9$, $K=0.02$ and $B=0$.
Figure~\ref{FIG1} shows the top views of relaxed compact bimeronium and bimeron structures. For the purpose of comparison, we also show the relaxed solutions of compact skyrmionium and skyrmion obtained with the same parameters but an easy-axis aligned along the $z$ axis.
The given skyrmionium with $Q=0$ [Fig.~\ref{FIG1}(e)] consists of an inner skyrmion with $Q=-1$ [Fig.~\ref{FIG1}(f)] and an outer skyrmion with $Q=+1$ [Fig.~\ref{FIG1}(g)].
Similarly, the corresponding bimeronium with $Q=0$ [Fig.~\ref{FIG1}(b)] exists as a combination of an inner bimeron with $Q=-1$ [Fig.~\ref{FIG1}(c)] and an outer bimeron with $Q=+1$ [Fig.~\ref{FIG1}(d)].
Namely, the bimeronium in in-plane magnetized magnets can be seen as a topological counterpart of the skyrmionium in out-of-plane magnetized magnets.

The total energy as well as different energy contributions for a relaxed bimeronium are given in Fig.~\ref{FIG2}(a). It shows that the competition among the FM NN, AFM NNN, and AFM NNNN exchange interactions is considerable. The magnetic anisotropy energy is positive, which means larger anisotropy constant could raise the total energy of a bimeronium and may reduce its stability.
A controllable degree of freedom of topological spin textures in frustrated magnetic systems is the helicity $\eta$~\cite{Leonov_NCOMMS2015,Lin_PRB2016A,Xichao_NCOMMS2017,Xia_PRApplied2019,Zhang_PRB2020,Xia_APL2020}.
It is found that the energy [Fig.~\ref{FIG2}(b)] and spin components [Fig.~\ref{FIG2}(c)] of the bimeronium are independent of its helicity.
We note that although the bimeronium helicity can freely vary between $0$ and $2\pi$, the orientation of the background spins outside the bimeronium is aligned with the easy axis orientation, i.e., the $x$ direction in this work.

As the magnetic anisotropy can be adjusted experimentally, we study the bimeronium structure for different anisotropy constants $K$, as shown in Fig.~\ref{FIG3}.
By increasing $K$ from $0$ to $0.095$, while keeping the easy-axis orientation aligned along the $x$ direction, the size of relaxed bimeronium decreases obviously. The spin component $m_x$ reduces with increasing $K$, while $m_y$ and $m_z$ do not depend on $K$ [Fig.~\ref{FIG3}(a)].
The total energy [Fig.~\ref{FIG3}(b)], anisotropy energy [Fig.~\ref{FIG3}(c)], AFM NNN exchange energy [Fig.~\ref{FIG3}(e)], and AFM NNNN exchange energy [Fig.~\ref{FIG3}(f)] increase with $K$, while the FM NN exchange energy [Fig.~\ref{FIG3}(d)] decreases with increasing $K$.
When $K$ is larger than certain threshold (i.e., $0.1$ in this work), the bimeronium structure becomes unstable and cannot exist in the system.

We also study the effect of an external in-plane magnetic field on the bimeronium structure (see \blue{supplemental material}).
As the bimeronium size is related to its spin component along the easy-axis orientation, which is $m_x$ in this work [Fig.~\ref{FIG3}(a)], we apply a magnetic field along the $x$ direction with a strength of $B_x$. Within a reasonable range of $B_x$ that does not destroy the bimeronium (i.e., in this work $B_x/1000=-0.2\sim 0.8$ in units of $J_{1}/a^{3}M_{\text{S}}=1$), it is found that the bimeronium size is insensitive to $B_x$.
The spin component $m_x$ increases with $B_x$, while $m_y$ and $m_z$ are independent of $B_x$.
The total energy, anisotropy energy, and FM NN exchange energy are proportional to $B_x$, while the AFM NNN and NNNN exchange energies are inversely proportional to $B_x$.
Note that a larger external magnetic field may lead to the deformation and annihilation of the bimeronium texture.

\begin{figure}[t]
\centerline{\includegraphics[width=0.479\textwidth]{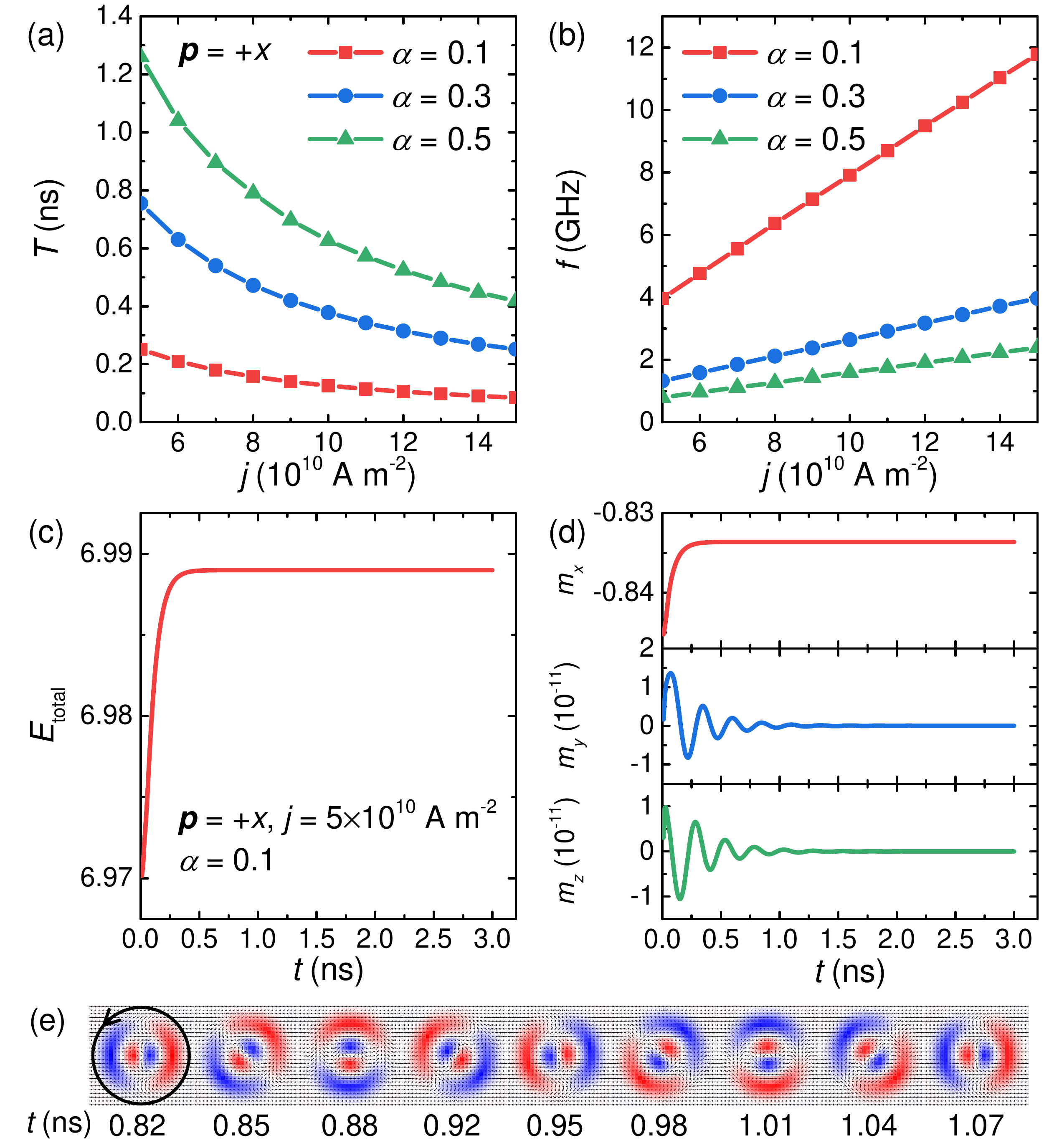}}
\caption{%
(a) Bimeronium rotation period $T$ as a function of driving current density $j$ for different damping parameters $\alpha$. The spin polarization direction $\boldsymbol{p}=+\hat{x}$.
(b) Bimeronium rotation frequency $f$ as a function of $j$ for different $\alpha$.
(c) Time-dependent total energy of a typical rotating bimeronium.
(d) Time-dependent spin components of a typical rotating bimeronium.
(e) Top views of a typical rotating bimeronium at selected times. The arrows represent the spin directions. The out-of-plane spin component ($m_z$) is color coded.
Here, $J_2=-0.8$, $J_3=-0.9$, $K=0.02$, and $B=0$. The initial bimeronium state consists of an inner bimeron with $Q=-1$ and an outer bimeron with $Q=+1$.
}
\label{FIG4}
\end{figure}

As shown in Fig.~\ref{FIG4}, we continue by investigating the dynamics of an isolated bimeronium driven by the damping-like SOT $\tau_\text{d}$.
We assume that $\tau_\text{d}$ is generated by the spin Hall effect in a heavy-metal substrate layer underneath the magnetic monolayer~\cite{Finocchio_JPD2016,Kang_PIEEE2016,Wanjun_PHYSREP2017,Fert_NATREVMAT2017,ES_JAP2018,Zhang_JPCM2020,Ado_PRB2017}.
We first consider that the direction of spin polarization is aligned along the easy-axis direction, i.e., $\boldsymbol{p}=+\hat{x}$.
A recent report~\cite{Zhang_PRB2020} numerically demonstrates that a frustrated bimeron shows SOT-induced self rotation when the spin polarization direction is parallel to the easy-axis direction.
Similarly, we find that the bimeronium also shows steady self rotation induced by the damping-like SOT when $\boldsymbol{p}=+\hat{x}$, which can be described by a theoretical approach (see \blue{supplemental material}).
The rotation period decreases with increasing driving current density [Fig.~\ref{FIG4}(a)] and the rotation frequency is proportional to the driving current density [Fig.~\ref{FIG4}(b)].
At a given current density, the inner and outer consisting bimerons of the bimeronium rotate in an identical manner, and the rotation frequency of the whole structure decreases with increasing damping parameter $\alpha$.
For the rotating bimeronium, its total energy [Fig.~\ref{FIG4}(c)] and spin components [Fig.~\ref{FIG4}(d)] are independent of time when the steady rotation state is reached.
Note that the rotating bimeronium is a dynamic stable object, which will not be annihilated even at a high speed rotation as the the inner and outer consisting bimerons rotate in an identical frequency and direction.
On the other hand, due to the non-circular symmetric out-of-plane spin structure of the bimerionium, its rotation can be observed by imaging the out-of-plane cores, while the self rotation of skyrmions and skyrmioniums can only be observed by imaging their in-plane spin components. Thus, we point out that it is possible to observe rotating bimeroniums by using the Kerr microscope system.

The bimeronium rotation depends on both the internal structure of the bimeronium as well as the spin polarization direction.
As shown in Fig.~\ref{FIG4}(e), when the bimeronium consists of an inner bimeron with $Q=-1$ and an outer bimeron with $Q=+1$, it shows counterclockwise rotation driven by the damping-like SOT with $\boldsymbol{p}=+\hat{x}$.
If the initial bimeronium structure consists of an inner bimeron with $Q=+1$ and an outer bimeron with $Q=-1$, it may show clockwise rotation driven by the damping-like SOT with $\boldsymbol{p}=-\hat{x}$ (see \blue{supplemental material}).

%%%%%%%%%%%%%%%%%%%%%%%%%%%%%%%%%%%%%%%%%%%%%%%%%%%%%%%%%%%%
%%%%%%%%%%%%%%%%%%%%%%%%%%%%%%%%%%%%%%%%%%%%%%%%%%%%%%%%%%%%

%%%%%%%%%%%%%%%%%%%%%%%%%%%%%%%%%%%%%%%%%%%%%%%%%%%%%%%%%%%%
\begin{figure}[t]
\centerline{\includegraphics[width=0.479\textwidth]{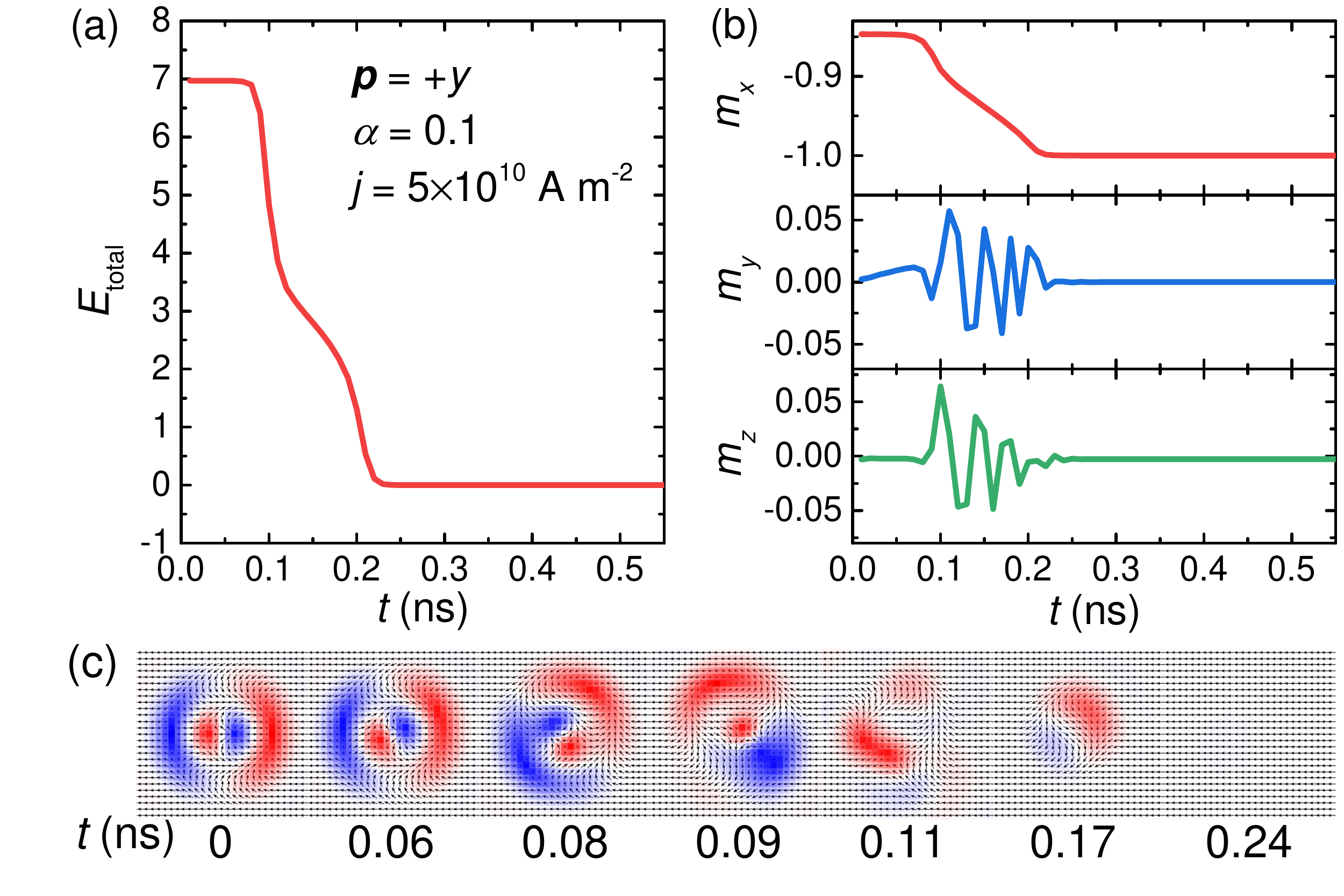}}
\caption{%
(a) Total energy of a bimeronium as a function of time during its annihilation induced by the SOT. The spin polarization direction $\boldsymbol{p}=+\hat{y}$.
(b) Spin components of a bimeronium as a function of time during its annihilation induced by the SOT.
(c) Top views of the bimeronium annihilation at selected times. The arrows represent the spin directions. The out-of-plane spin component ($m_z$) is color coded.
Here, $J_2=-0.8$, $J_3=-0.9$, $K=0.02$, and $B=0$. The initial bimeronium state consists of an inner bimeron with $Q=-1$ and an outer bimeron with $Q=+1$.
}
\label{FIG5}
\end{figure}
%%%%%%%%%%%%%%%%%%%%%%%%%%%%%%%%%%%%%%%%%%%%%%%%%%%%%%%%%%%%

As reported in Ref.~\onlinecite{Zhang_PRB2020}, a frustrated bimeron could show SOT-induced translational motion when the spin polarization direction is perpendicular to the easy-axis direction.
However, we find that the frustrated bimeronium cannot be driven into steady motion when the spin polarization direction is perpendicular to the easy-axis direction, i.e., $\boldsymbol{p}=\pm\hat{y}$.
Instead, the damping-like SOT leads to the deformation and annihilation of the bimeronium structure, as shown in Fig.~\ref{FIG5}.
To be more specific, the damping-like SOT leads to different rotation behaviors of the inner and outer consisting bimerons of the bimeronium, and the bimeronium structure is twisted upon the application of the SOT. When the out-of-plane cores of the inner and outer consisting bimerons merge together, the bimeronium is thus destroyed and annihilated.
The total energy [Fig.~\ref{FIG5}(a)] and spin component $m_x$ [Fig.~\ref{FIG5}(b)] decrease significantly during the SOT-induced annihilation of the bimeronium.
The spin components $m_y$ and $m_z$ also show certain fluctuations during the annihilation process.
Note that the initial bimeronium structure in Fig.~\ref{FIG5} consists of an inner bimeron with $Q=-1$ and an outer bimeron with $Q=+1$. For the bimeronium structure consisting of an inner bimeron with $Q=+1$ and an outer bimeron with $Q=-1$, it also shows deformation and annihilation when the damping-like SOT with $\boldsymbol{p}=\pm\hat{y}$ is applied (see \blue{supplemental material}).

%-%-%-%-%-%-%-%-%-%-%-%-%-%-%-%-%-%-%-%-%-%-%-%-%-%-%-%-%-%-%
%\section{Conclusion}
%\label{se:Conclusion}
%-%-%-%-%-%-%-%-%-%-%-%-%-%-%-%-%-%-%-%-%-%-%-%-%-%-%-%-%-%-%

In conclusion, we have studied the static structures and SOT-induced dynamics of an isolated bimeronium in a frustrated magnetic monolayer with competing FM and AFM exchange interactions.
Note that the small FM NN and large AFM NNN exchange interactions could be realized in low-dimensional compound Pb$_2$VO(PO$_4$)$_2$ with frustrated square lattice~\cite{Kaul_2004}.
The bimeronium structure carries a topological charge of $Q=0$ but it can be regarded as a combination of two bimerons with opposite topological charges. Namely, it may consist of an inner bimeron with $Q=-1$ and an outer bimeron with $Q=+1$, and it may also consist of an inner bimeron with $Q=+1$ and an outer bimeron with $Q=-1$.

It is found that the frustrated bimeronium energy is independent of its helicity in the in-plane magnetized system, however, the size and energy of a bimeronium is subject to the easy-axis magnetic anisotropy. A larger anisotropy will lead to a smaller compact bimeronium with higher total energy. Indeed, extremely large anisotropy may result in the instability of the bimeronium structure.
On the other hand, the bimeronium energy can be subtly adjusted by an external in-plane magnetic field, however, the bimeronium size is insensitive to the magnetic field.

In this work, we have also numerically demonstrated that the bimeronium can be driven into steady rotation by the damping-like SOT, where the spin polarization direction is parallel to the easy-axis direction.
It is found that the rotational dynamics depends on both the internal bimeronium structure and the spin polarization direction. In particular, when the spin polarization direction is perpendicular to the easy-axis, the bimeronium is annihilated by the damping-like SOT.

We point out that the rotational feature of a bimeronium may be used for building a multi-state memory device based on bimeronium-hosting nanodots~\cite{Hou_NC2020}, where a bimeronium with variable helicity values in a unit nanodot can be used to store different information.
The current-controlled rotation of a bimeronium could also be useful for future spin-wave applications, where arrays of bimeroniums serve as reconfigurable spin wave guides.
We believe our results are important for understanding the frustrated bimeronium structures, and can provide guidelines for the design of spintronic devices based on bimeroniums.

\vbox{}

\textsl{See supplemental material for additional simulation results.}

\vbox{}

%-%-%-%-%-%-%-%-%-%-%-%-%-%-%-%-%-%-%-%-%-%-%-%-%-%-%-%-%-%-%
%\begin{acknowledgments}
%\end{acknowledgments}
%-%-%-%-%-%-%-%-%-%-%-%-%-%-%-%-%-%-%-%-%-%-%-%-%-%-%-%-%-%-%

This work was primarily supported by the National Natural Science Foundation of China (Grant No.~12004320). X.Z. also acknowledges the support by the Guangdong Basic and Applied Basic Research Foundation (Grant No.~2019A1515110713).
M.E. acknowledges the support by the Grants-in-Aid for Scientific Research from JSPS KAKENHI (Grant Nos.~JP18H03676 and JP17K05490) and the support by CREST, JST (Grant Nos.~JPMJCR20T2 and JPMJCR16F1).
O.A.T. acknowledges the support by the Australian Research Council (Grant No.~DP200101027), the Cooperative Research Project Program at the Research Institute of Electrical Communication, Tohoku University (Japan), and by the NCMAS grant.
G.Z. acknowledges the support by the National Natural Science Foundation of China (Grant Nos.~51772004, 51771127, and 51571126).
X.L. acknowledges the support by the Grants-in-Aid for Scientific Research from JSPS KAKENHI (Grant Nos.~JP20F20363).
Y.Z. acknowledges the support by the Guangdong Special Support Project (Grant No.~2019BT02X030), Shenzhen Peacock Group Plan (Grant No.~KQTD20180413181702403), Pearl River Recruitment Program of Talents (Grant No.~2017GC010293), and National Natural Science Foundation of China (Grant Nos.~11974298 and 61961136006).

\vbox{}
\noindent
\textbf{DATA AVAILABILITY}

The data that support the findings of this study are available from the corresponding author upon reasonable request.

%-%-%-%-%-%-%-%-%-%-%-%-%-%-%-%-%-%-%-%-%-%-%-%-%-%-%-%-%-%-%

%-%-%-%-%-%-%-%-%-%-%-%-%-%-%-%-%-%-%-%-%-%-%-%-%-%-%-%-%-%-%

%%%%%%%%%%%%%%%%%%%%%%%%%%%%%%%%%%%%%%%%%%%%%%%%%%%%%%%%%%%%
\end{document}